# Hydration Features on Near-Earth Objects: Integrating New Data with Prior Results

**L. E. McGraw[1], C. A. Thomas[1], J. P. Emery[1], A. S. Rivkin[2]**

[1]Northern Arizona University, Department of Astronomy and Planetary Science, P.O. Box 6010, Flagstaff, AZ 86011, USA (lem366@nau.edu)

[2]JHU/APL, 211100 Johns Hopkins Road, Laurel, MD 20723

**Abstract**

Near-Earth objects (NEOs) are excellent laboratories for testing processes that affect airless bodies, as well as informing us about Solar System history. Though most NEOs are nominally anhydrous because they formed inside the Solar System frost line and their surface temperatures are high enough to remove volatiles, a 3-μm feature typically indicative of $OH/H_2O$ has been identified on several such bodies. Possible sources for $OH/H_2O$ on these bodies include carbonaceous chondrite impactors or interactions with protons implanted by solar wind. The MIT-Hawaii Near-Earth Object Spectroscopic Survey (MITHNEOS) began its 3-μm observation campaign of NEOs in 2022 and has obtained spectral data of 15 predominantly nominally anhydrous (i.e., mostly S-complex or V-type) targets using NASA's Infrared Telescope Facility's (IRTF) near-infrared spectrometer, SpeX. Spectra were collected using both prism (0.7-2.52 μm) and LXD_short (1.67-4.2 μm) modes to accurately characterize asteroid spectral type and the 3-μm region. Four of the 15 NEOs observed exhibit a 3-μm feature, exhibiting band shapes similar to those identified in a previous NEO survey (McGraw et al. 2022), which found a trend between hydration band presence and large aphelion (i.e., Q > 2.06 AU). Combining our new observations with the pre-existing database of NEO 2-4-μm data revealed that band depth increases with decreasing orbital inclination and that all NEOs with hydration bands have i < 27° with most having i < 14°. We find that NEOs with low inclination and large aphelia are the most likely bodies in near-Earth space to possess surficial $OH/H_2O$.

# 1 Introduction

Asteroids are the relatively unaltered remnants of the solar nebula, so can yield information about Solar System formation and evolution. They exhibit a wide range of sizes, surface temperatures, and compositions, from silicaceous to carbonaceous to metallic. Although most orbit between Mars and Jupiter, gravitational and non-gravitational forces can change asteroidal orbits, particularly for smaller objects, so that they cross planetary orbits and potentially become impactors. Near-Earth objects (NEOs) are a subset of the overall asteroid population with perihelia < 1.3 AU, that are fragments of main-belt asteroids. Because they are derived from main-belt asteroids, they exhibit a wide range of physical characteristics. Their close proximity to Earth makes them more accessible by spacecraft and those that impact Earth are the cosmic progenitors of meteorites (e.g., Thomas et al. 2014). NEOs are potentially valuable as sources of various resources such as volatiles for fuel and materials rare on Earth or expensive to ship into space for use there (e.g., Elvis 2014). Additionally, some NEOs are classified as Potentially Hazardous Asteroids (PHAs), and impact mitigation is aided by compositional analyses and general NEO studies. Their wide range of characteristics makes NEOs excellent laboratories for a variety of investigations, as many factors can be explored by studying multiple objects.

Most airless inner Solar System objects are nominally anhydrous, as most have relatively high surface temperatures and formed inside the frost line, where water ice is unstable in a vacuum (e.g., Grossman & Larimer 1974; Muralidharan et al. 2008). However, numerous recent studies have shown that non-carbonaceous inner Solar System bodies, such as the Moon, the Main Belt asteroid (MBA) (4) Vesta, numerous S-complex MBAs, and several NEOs, contain hydroxide and/or water ($OH/H_2O$) on their surfaces, as evidenced by a 3-μm spectroscopic absorption feature (e.g., Sunshine et al. 2009; Pieters et al. 2009; Clark 2009; Simon et al. 2019; Ruiz et al. 2020; Honniball et al. 2021; De Sanctis et al. 2012; Prettyman et al. 2012; McCord et al. 2012; Reddy et al. 2012**a**; De Sanctis et al. 2013; McAdam et al. 2024; Rivkin et al. 2018; McGraw et al. 2022; McGraw et al. 2024a). These detections show that hydrated material is far more common in near-Earth space than originally expected.

Four potential sources for inner Solar System (i.e., within Jupiter's 3:1 mean motion resonance at ~2.5 AU) $OH/H_2O$ exist: 1) native phyllosilicates, 2) exogenous cometary material, 3) exogenous carbonaceous material, and 4) the solar wind. An asteroid with native phyllosilicates, which contain OH within their mineral structure, would exhibit a deep 3-μm feature (band depth at 2.9 μm > 10%) that linearly increases in reflectance from the band minimum to the reflected continuum (e.g., Rivkin & Emery 2010; Takir & Emery 2012). Such minerals can also be formed via aqueous alteration of non-hydrated minerals, as seen on JAXA's Hayabusa2 target (162173) Ryugu and NASA's OSIRIS-REx target (101955) Bennu (e.g., Nakamura et al. 2023; Zega et al. 2025). These two objects, both of which are C-complex NEOs whose parent bodies likely contained water ice, exhibit shallower 3-μm features than expected for a body with native phyllosilicates. Spacecraft and lab results show Ryugu and Bennu possess 2.7 μm bands with depths between 7 and 18% (e.g., Kitazato et al. 2019; Hamilton et al. 2019; Simon et al. 2020; Pilorget et al. 2022, 2025). However, these depths at wavelengths observable from the ground (i.e., 2.9 μm) are closer to 5%, roughly half the depth expected for a band at this wavelength caused by native phyllosilicates. This volatile source is not likely for S-complex or V-type asteroids, which are nominally not expected to contain such minerals or components (Vilas & Gaffey 1989; Rivkin et al. 1995; Rivkin et al. 2000; Landsman et al. 2015; Takir et al. 2017). As most NEOs are spectrally identified as S-complex because they are primarily sourced from the inner Main Belt (e.g., Binzel et al. 2015; Marsset et al. 2022), phyllosilicates are an unlikely, though not impossible (e.g., Hutchison et al. 1987), source of $OH/H_2O$ on these bodies.

Exogeneous carbonaceous material impactors are the most likely source of Vesta's $OH/H_2O$, as the strongest absorption features are correlated with dark material found near impact craters (De Sanctis et al. 2012). Both this source and cometary impacts were considered improbable for most NEOs as impacts are rare in near-Earth space and cometary/carbonaceous material is mostly found in the outer Main Belt and beyond. However, this delivery mechanism is likely at work on the largest NEO, (1036) Ganymed and possibly other NEOs, as NEOs with aphelia in the Main Asteroid Belt seem to preferentially exhibit a 3-μm feature (Rivkin et al. 2018; McGraw et al. 2022; McGraw et al. 2024a). Exogeneous carbonaceous and cometary material contain water, potentially in addition to hydroxide, so causes spectra to exhibit shallow, bowl-like features in the 2-4-μm spectral region (e.g., Greenwood et al. 2011; Neumann et al. 2013; McCord et al. 2012; De Sanctis et al. 2012; Rivkin & Emery 2010; Takir & Emery 2012).

$OH/H_2O$ on the majority of NEOs is most likely produced, at least in part, by interactions with the solar wind. Lunar $OH/H_2O$ is hypothesized to have been delivered by the solar wind as well (e.g., Sunshine et al. 2009; Wöhler et al. 2017; Bandfield et al. 2018; Laferriere et al. 2022). High-speed protons in the solar wind bombard oxygen-rich surfaces, breaking pre-existing bonds and reforming with oxygen to form OH and/or $H_2O$ (Starukhina 2001). This process is expected to more easily create OH than $H_2O$ (e.g., Starukhina 2001, 2003, 2006; Farrell et al. 2015, 2017) so should create an absorption feature with a similar shape to that of native phyllosilicates but much shallower (Band Depth $\lesssim$ 5%; e.g., Sunshine et al. 2009). Unfortunately, the most definitive method to differentiate between these sources of OH is band center analysis. Due to atmospheric water vapor, ground-based observations cannot spectrally

"see" the center of the 3-µm band. The second largest NEO, (433) Eros, along with several others NEOs likely contain OH/$H_2O$ sourced by this mechanism, though a combination of processes may be present (Rivkin et al. 2018; McGraw et al. 2022; McGraw et al. 2024a).

The MIT-Hawaii Near-Earth Object Spectroscopic Survey (MITHNEOS) has observed over 1000 asteroids in the 0.7-2.5 µm region since its inception in 2004. The MITHNEOS project recently broadened its spectral range to include the 2-4-µm spectral region in order to study and characterize OH/$H_2O$ on NEO surfaces. Our new observations searching for 3-µm absorption features on NEOs builds on previous work characterizing OH/$H_2O$ on these bodies. We add our targets to the pre-existing database of NEOs with such spectra (e.g., Rivkin et al. 2018; McGraw et al. 2022; McGraw et al. 2024a; McGraw et al. 2024b) to further constrain and predict OH/$H_2O$ presence in near-Earth space, particularly on nominally anhydrous, siliceous bodies. We observed 15 NEOs from Fall 2022 to Spring 2025 using NASA's Infrared Telescope Facility (IRTF), increasing the number of NEOs with 2-4-µm spectra by ~50%.

## 2 Methods

### 2.1 Data Acquisition and Processing

All targets observed by this work used NASA's IRTF's near-infrared spectrograph SpeX (Table 1; Rayner et al. 2003). We used two modes on SpeX: long-wavelength cross-dispersed short mode (LXD_short; 1.67-4.2 µm) to ascertain the presence of a 3-µm absorption feature indicative of OH/$H_2O$ and prism mode (0.7-2.5 µm) to characterize each target's mineralogy. As in similar 3-µm NEO studies (e.g., Rivkin et al. 2018; McGraw et al. 2022; McGraw et al. 2024a), we used the 0.8x15" slit, resulting in resolving powers (R=$\lambda/\Delta\lambda$) of ~1000 and ~100 for LXD and prism modes, respectively, and spectra that were roughly oversampled by a factor of 8. These spectra were rebinned by a factor of 9 during processing to achieve the desired signal-to-noise of ~50. Following the procedure outlined in McGraw et al. (2022; 2024b), we used the MIT Optical Rapid Imaging System (MORIS) along with the IRTF's infrared camera GuideDog to track our targets and ensure appropriate placement within the slit. The telescope nodded within the slit for background sky emission subtraction, and we alternated between observing a local solar-like star and the asteroid every 10 and 30 minutes, respectively, for atmospheric and telluric corrections.

**Table 1:** Observing parameters and conditions of NEOs

| Object | Date (UT) | V mag[a] | LXD Mid-time (UT) | LXD Int. Time (s) x Images | Prism Mid-time (UT) | Prism Int. Time (s) x Images | Standard Star | Solar Dist. (AU)[a] | Earth Dist. (AU)[a] | Phase Angle[a] | Weather (Seeing) |
|---|---|---|---|---|---|---|---|---|---|---|---|
| 887 Alinda | 8-Jan-2025 | 9.7 | 10:08 | 30 x 62 | 8:50 | 21x8;15x4 | HD 42380 | 1.063 | 0.082 | 14.85 | Clear, humid (0.8") |
| | 10-Jan-2025 | 9.5 | 11:30 | 30 x 48 | 9:38 | 21 x 16 | HD 45395 | 1.065 | 0.083 | 11.01 | Cloudy, foggy |
| | 12-Jan-2025 | 9.4 | 11:54 | 30 x 58 | 12:50 | 21 x 4 | HD 50692 | 1.067 | 0.085 | 8.72 | Mostly clear (1.13") |
| 1685 Toro | 28-Jan-2024 | 12.7 | 6:51 | 30 x 100 | 5:11 | 30 x 8 | HD 23066 | 1.024 | 0.149 | 70.50 | Clear, windy (0.649") |
| 4954 Eric | 7-Feb-2025 | 13.7 | 8:25 | 30 x 202 | -- | -- | HD 33442 | 1.375 | 0.600 | 39.23 | Clear (0.64") |
| 161989 Cacus | 3-Sep-2022 | 13.5 | 14:20 | 30 x 96 | 12:17 | 30x16;40x8 | HD 20006 | 1.035 | 0.061 | 62.70 | Clear (0.56") |
| | 4-Sep-2022 | 13.6 | 13:17 | 30 x 134 | -- | -- | HD 20101 | 1.038 | 0.063 | 60.31 | Clear (0.64") |
| 164121 2003 YT1 | 4-Nov-2023 | 12.2 | 6:50 | 30 x 128 | 9:02 | 30 x 8 | HD 214999 | 1.021 | 0.060 | 59.66 | Clear (0.55") |
| 187026 2005 EK70 | 18-Feb-2024 | 12.9 | 11:42 | 30 x 240 | 15:11 | 40 x 12 | HD 89277 | 1.086 | 0.100 | 4.70 | Clear, windy (1.72") |
| 363027 1998 ST27 | 10-Oct-2024 | 13 | 8:10 | 30 x 192 | 5:08 | 60 x 8 | HD 222612 | 1.026 | 0.029 | 21.13 | Cirrus (0.715") |
| 415009 2011 UL21 | 29-Jun-2024 | 11.7 | 6:51 | 30 x 52 | -- | -- | HD 122607 | 1.039 | 0.049 | 61.52 | Clear, humid (0.282") |
| | 10-Jul-2024 | 14.3 | -- | -- | 9:28 | 90x8;120x8 | Land. 110-361 | 1.151 | 0.194 | 42.59 | Clear, humid, windy (2.5") |
| 424482 2008 DG5 | 1-Jun-2025 | 13.7 | 8:55 | 30 x 184 | 5:33 | 40 x 24 | HD 124244 | 1.034 | 0.029 | 45.89 | Mostly clear (0.8") |
| 439437 2013 NK4 | 17-Apr-2024 | 12.6 | 13:22 | 30 x 160 | -- | -- | HD 138442 | 1.029 | 0.029 | 29.47 | Clear (0.7") |
| | 21-Apr-2024 | 13.4 | -- | -- | 6:35 | 40 x 32 | Land. 105-56 | 1.063 | 0.058 | 7.70 | Clear (0.669") |
| 458122 2010 EW45 | 22-Dec-2024 | 13.7 | 7:10 | 30 x 176 | 10:04 | 60 x 12 | SAO 129922 | 1.015 | 0.063 | 58.68 | Clear (0.82") |
| 756998 2024 CR9 | 9-Jun-2024 | 14.1 | 11:48 | 30 x 130 | -- | -- | HD 170623 | 1.062 | 0.050 | 19.65 | Clear (0.65") |
| | 8-Aug-2024 | 18.1 | -- | -- | 13:13 | 120 x 24 | Land. 133-276 | 1.238 | 0.252 | 24.56 | Clear (0.582") |
| 1998 HH49 | 19-Oct-2023 | 14 | 8:05 | 30 x 168 | 5:23 | 30x8;40x16 | SAO 91793 | 1.016 | 0.021 | 18.98 | Clear (0.64") |
| 2006 WB | 27-Nov-2024 | 13.7 | 8:41 | 30 x 192 | 11:34 | 40 x 16 | HD 7354 | 0.991 | 0.006 | 42.74 | Clear (0.76") |
| 2024 ON | 16-Sep-2024 | 12.7 | 6:52 | 30 x 88 | -- | -- | HD 176247 | 1.008 | 0.009 | 70.91 | Clear, humid (0.55") |
| | 7-Sep-2024 | 16.7 | -- | -- | 6:06 | 120x16 | Land. 110-361 | 1.021 | 0.052 | 74.26 | Mostly clear (0.86") |

[a]From JPL Horizons

Again following the methods used in previous 3-μm NEO surveys, we used the Spextool IDL-based software package to extract, telurically correct, combine, and merge the LXD data into one continuous spectrum for each target (Cushing et al. 2004; Vacca et al. 2003). The prism data was also extracted and combined using the same package but was atmospherically corrected using an ATRAN-based IDL code (Lord 1992; MacLennan et al. 2024b). All prism and LXD data were normalized to 1 at 2.2 μm. By using the same data processing methods as the above-referenced NEO surveys, we can easily add our newly acquired data to the overall NEO 3-μm dataset thereby increasing the overall sample size available for OH/$H_2O$ characterization analysis.

## 2.2 Thermal Tail Removal

Due to the proximity of our targets to the Sun during our observations and thus high surface temperatures, all processed asteroid spectra exhibit a significant amount of emitted flux (a "thermal tail") beginning roughly around 2.8 μm, with the exact wavelength dependent on each target's surface temperature. As we are only interested in the component of each spectrum caused by reflected light, this thermal tail must be removed before analyzing each spectrum for any absorption features. Our thermal model is based on the Near Earth Asteroid Thermal Model (NEATM; Harris 1998), which applies assumptions applicable to NEOs to the Standard Thermal Model (STM). These models use subsolar temperature to determine the distribution of surface temperatures and calculate thermal flux (Lebofsky & Spencer 1989), which is then subtracted from the full spectrum to produce a reflectance-only spectrum.

Atmospheric telluric contamination between 2.45 μm and 2.85 μm causes near total signal loss in this spectral region, so we remove all data from this range. Based on laboratory spectra of analog materials, we assume the asteroid reflectance spectra to be linear and featureless beyond 3.3 μm (e.g., McGraw et al. 2022), so the reflected continuum is linearly extrapolated from 2.45 μm, which is also the red edge of the 2-μm band on S-complex asteroids, to the end of the spectrum at 4.0 μm. The best-fit thermal model is found when the longer wavelength data lies along the continuum, which is achieved by iteratively varying the beaming parameter, continuum slope, and continuum vertical position. The difference between the continuum and reflectance at 2.9 μm is the band depth (Table 3).

## 2.3 Band Parameter Analysis

### 2.3.1 Prism Bands

We conducted band parameter analysis on the 1- and 2-μm absorption bands (BI and BII, respectively) that are indicative of pyroxene and olivine composition for the S-complex and V-type asteroids on our target list (Table 2). We used an IDL-based routine called the Spectral Analysis for Asteroid Reflectance Investigation (SAARI; MacLennan et al. 2024a) to calculate the band center wavelengths (BIC and BIIC) and the ratio of the area of BII to that of BI (band-area ratio or BAR) and the associated uncertainties. However, the SAARI-calculated uncertainties do not take the spectral resolution of SpeX into account (R~200). We therefore assumed the error in band center is the minimum band width of SpeX for BI and BII, unless that calculated by SAARI was larger. The uncertainty in BAR was propagated from the chosen BI and BII wavelength uncertainties, except in cases in which the SAARI-calculated error was larger. We then calculated pyroxene and olivine mineralogies, i.e., mole percent (mol%) ferrosilite (Fs) and fayalite (Fa), respectively, as well as the ratio of olivine to the olivine plus pyroxene content for the S-complex asteroids using BIC, BIIC, and BAR (Sanchez et al. 2020). For the V-type asteroids, we used BIC and BIIC to calculate ferrosilite (iron pyroxene end member) and wollastonite (Wo; calcium pyroxene end member), as well as the ratio of magnesium to the total magnesium and ferrous iron content (Burbine et al. 2007). All mineralogical calculations done in this study take numerous factors such as surface temperature (e.g., Sanchez et al. 2012; Reddy et al. 2012b), signal-to-noise ratio (S/N; Sanchez et al. 2020), phase angle (e.g., Sanchez et al. 2012; Reddy et al. 2012b), and instrument wavelength range (e.g., Lindsay et al. 2016; Sanchez et al. 2020) into account by applying any necessary corrections where appropriate.

**Table 2**: Physical and orbital characteristics of NEOs

| Object | D (km) | Spectral Type | q (AU) | Q (AU) | i (°) |
|---|---|---|---|---|---|
| 887 Alinda | 4.2 | S | 1.07 | 3.89 | 9.39 |
| 1685 Toro | 3.4 | Sq | 0.77 | 1.96 | 9.38 |
| 4954 Eric | 10.8 | Sr | 1.10 | 2.90 | 17.43 |
| 161989 Cacus | 1.9 | Q | 0.88 | 1.36 | 26.06 |
| 164121 2003 YT1 | 1.717 | V | 0.79 | 1.43 | 44.06 |
| 187026 2005 EK70 | 0.812 | Q | 0.83 | 1.09 | 30.00 |
| 363027 1998 ST27 | 0.578 | C/X | 0.39 | 1.25 | 21.06 |
| 415029 2011 UL21 | 2.3 | S | 0.74 | 3.51 | 34.85 |
| 424482 2008 DG5 | 0.41 | Sq | 0.95 | 1.56 | 5.71 |
| 439437 2013 NK4 | 0.59 | V | 0.46 | 1.59 | 6.52 |
| 458122 2010 EW45 | 0.759 | Q | 0.68 | 3.44 | 2.11 |
| 756998 2024 CR9 | 0.54 | S | 1.06 | 3.65 | 4.36 |
| 1998 HH49 | 0.19 | Q | 0.77 | 2.33 | 8.42 |
| 2006 WB | 0.1 | Xc | 0.70 | 1.00 | 4.87 |
| 2024 ON | 0.355 | Sr | 1.01 | 3.73 | 7.74 |

### 2.3.2 OH/H₂O Band

After the thermal tail is removed, the OH/$H_2O$ band depth at a specified wavelength ($BD_\lambda$) is calculated as a percent by subtracting the reflectance of the spectrum at the specified wavelength ($RB_\lambda$) from the reflectance of the continuum at the same wavelength ($RC_\lambda$) and dividing it by the reflectance at the continuum ($BD = \frac{RC_\lambda - RB_\lambda}{RC_\lambda} * 100$). The primary band depth is calculated at 2.9 μm, the shortest wavelength with a reliable reflectance value that is closest to the band center around 2.7 μm (see Section 1) and outside the spectral region of total signal loss due to atmospheric water vapor. We also calculated the band depths at 2.95 μm, 3.0 μm, 3.05 μm, and 3.1 μm to create a simplified band shape used to compare with other datasets (e.g., McGraw et al. 2022). For each depth calculated, $RB_\lambda$ was determined by taking the robust mean of the data points in a 0.05-μm band width centered on the wavelength of interest. We calculated the error in band depth by assessing the point-to-point spread in the data and the subjectivity of continuum slope and placement using the method described in previous NEO surveys (McGraw et al. 2022, 2024a), which includes using standard error propagation techniques and adding in quadrature the error calculated from varying the chosen reflected continuum. The reported band depth errors in Table 3 are 1σ, and any spectrum with a band depth greater than the reported error is considered to potentially have an OH/$H_2O$ feature; bands deeper than the 2σ-error are considered definitively present.

## 3 Results

We observed 15 NEOs as part of MITHNEOS' campaign to further characterize OH/$H_2O$ in near-Earth space (Table 1) and collected prism and LXD data for all but (4954)

Eric; we used prism data collected during a previous MITHNEOS campaign for this target[1] for the band parameter analysis. All but one of these targets ((1685) Toro; McGraw et al. 2022) has never been observed in the 2-4 µm spectral region before this campaign. One target, (887) Alinda, was observed three times to conduct a rotationally resolved spectral search for a 3-µm feature. Alinda was the only NEO observed during our campaign that achieved sufficient brightness (V mag < 10) to conduct such a study. See Appendix A for all spectra collected during this campaign. We used the online Bus-DeMeo classification tool (DeMeo et al. 2009) to determine spectral type. One of the 15 targets exhibits a definitive (BD > 2σ) 3-µm feature and three exhibit a potential (BD > 1σ) feature. A histogram of S/N calculated at $2.9 \pm 0.025$ µm, the band center and width at and over which the primary depth is calculated, shows a bimodal distribution. The approximate S/N minimum between the two peaks is 14, which we used as a cutoff to determine which of the remaining 12 targets have spectra of sufficient S/N to definitively state that they do not exhibit a 3-µm feature (e.g., McGraw et al. 2022). Using this cutoff, spectra of three asteroids have too low S/N (i.e., S/N < 14) to determine the presence or earth of a feature and nine definitively lack a 3-µm absorption feature. Prism spectra are shown in Appendix A.

### 3.1 Detections

Four NEOs were found to have a 3-µm feature: (161989) Cacus, (756998) 2024 CR9, 1998 HH49, and 2006 WB (Figure 1). Cacus, 2024 CR9, and 1998 HH49 exhibit potential absorption features indicative of OH/$H_2O$ are all in the S-Complex. Cacus is the largest of the four asteroids (diameter = 1.9 km) with 3-µm features and has a band depth of $3.3 \pm 2.4\%$. 2024 CR9 and 1998 HH49 have band depths of $11.7 \pm 8.6\%$ and $17.4 \pm 10.9\%$, respectively. 2006 WB's prism spectrum most closely resembles an Xc-type asteroid and is one of only two non-silicaceous (i.e., not S-complex or V-type) NEOs observed (see Section 3.2). It is also the only NEO observed in this campaign with a definitive 3-µm feature with a depth of $9.9 \pm 3.4\%$. These four NEOs exhibit three different 3-µm band shapes, with 2024 CR9 and 1998 HH49 exhibiting similarly shaped features. Prism spectra for these four NEOs are shown in Appendix A.

**Table 3:** NEO Band Depths

| Object | | 3-µm Band? | Band Depth (w/ 1σ errors) |
|---|---|---|---|
| 887 | Alinda (8-Jan) | No | $0.4 \pm 1.9\%$ |
| 887 | Alinda (10-Jan) | No* | $12.7 \pm 3.8\%$ |
| 887 | Alinda (12-Jan) | No | $0.9 \pm 5.9\%$ |
| 1685 | Toro | No | $1.8 \pm 2.7\%$ |
| 4954 | Eric | No | $1.4 \pm 4.2\%$ |
| 161989 | Cacus | Yes (1σ) | $3.3 \pm 2.4\%$ |
| 164121 | 2003 YT1 | No | $-0.3 \pm 3.2\%$ |
| 187026 | 2005 EK70 | No | $-1.3 \pm 4.9\%$ |
| 363027 | 1998 ST27 | No | $-0.4 \pm 3.1\%$ |
| 415029 | 2011 UL21 | No | $-3.9 \pm 3.6\%$ |
| *424482* | *2008 DG5*[†] | -- | *$4.2 \pm 8.9\%$* |
| 439437 | 2013 NK4 | No | $0.1 \pm 1.6\%$ |
| *458122* | *2010 EW45*[†] | -- | *$0.7 \pm 9.8\%$* |
| 756998 | 2024 CR9 | Yes (1σ) | $11.7 \pm 8.6\%$ |
| | 1998 HH49 | Yes (1σ) | $17.4 \pm 10.9\%$ |
| | 2006 WB | Yes (2σ) | $9.9 \pm 3.4\%$ |
| | *2024 ON*[†] | -- | *$3.0 \pm 5.8\%$* |

*\*See text for discussion of this result.*
*[†]Italicized data represent low S/N spectra.*

## 3.2 Non-Detections

Eight of the remaining eleven NEOs without a 3-μm feature have spectra of sufficiently high S/N to definitively conclude they do not possess a band with a depth of a few % or greater (Table 3; Figure 2). The spectra for the other three NEOs ((424482) 2008 DG5, (458122) 2010 EW45, and 2024 ON) are too noisy to confirm the presence or dearth of a feature in the 2-4 μm spectral region (Figure 3). (887) Alinda was observed three times to search for hydration band depth variations across its surface. Though one spectrum showed a band depth greater than 2σ, the presence of significant atmospheric contamination present in the 2-μm band even after the application of telluric corrections limit the trustworthiness of the spectrum (Figure 2). The other two Alinda spectra do not show a band within the level of the noise and both are considered high S/N spectra. Alinda will be discussed in more detail in Section 5. The spectrum we collected of (1685) Toro did not exhibit a band, which agrees with the spectrum published by McGraw et al. (2022) for this object. Prism spectra for these eleven NEOs are shown in Appendix A.

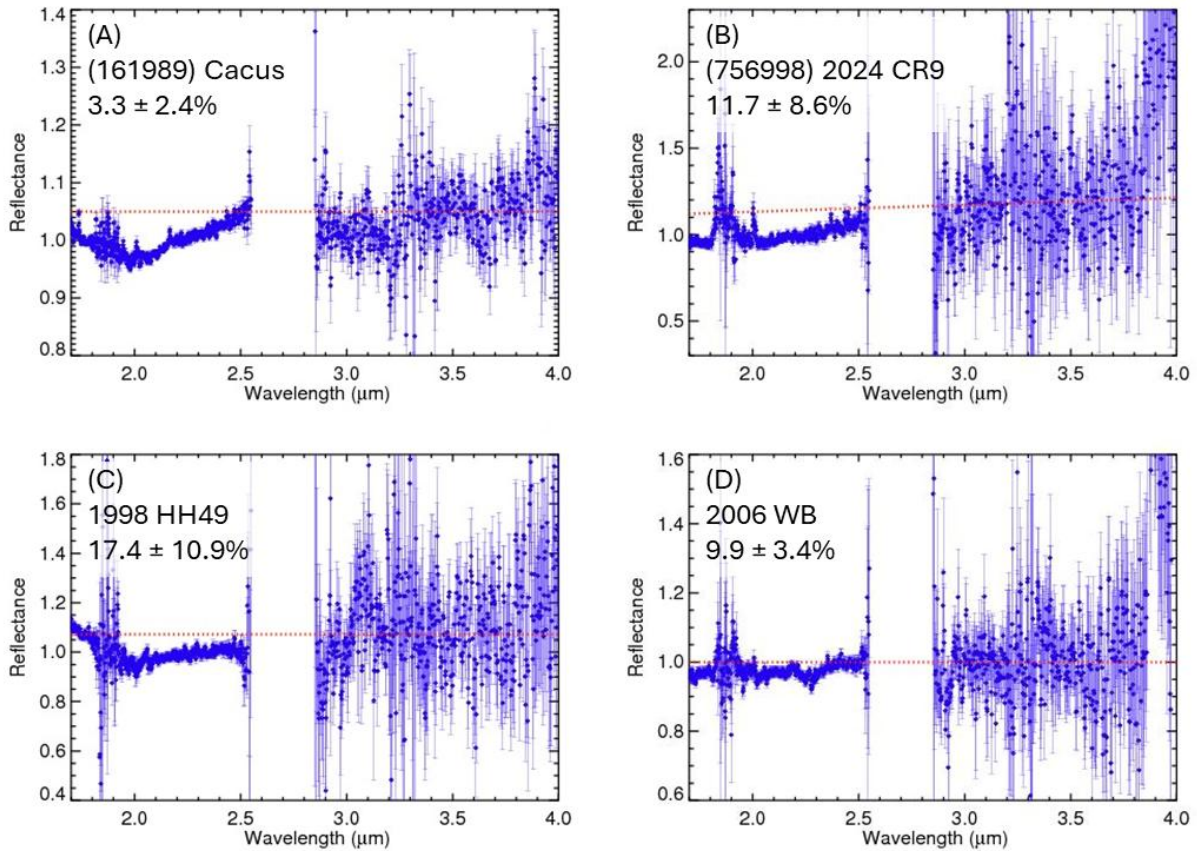

**Figure 1:** Spectra of NEOs with (potential) 3-μm absorption features. (A) Spectrum of (161989) Cacus as observed on 3&4 September 2022. (B) (756998) 2024 CR9 as observed on 9 June 2024. (C) 1998 HH49 as observed on 19 October 2023. (D) 2006 WB as observed on 27 November 2024. The blue points represent the reflectance spectra (thermal component removed) and the red dashed lines are the reflected continua.

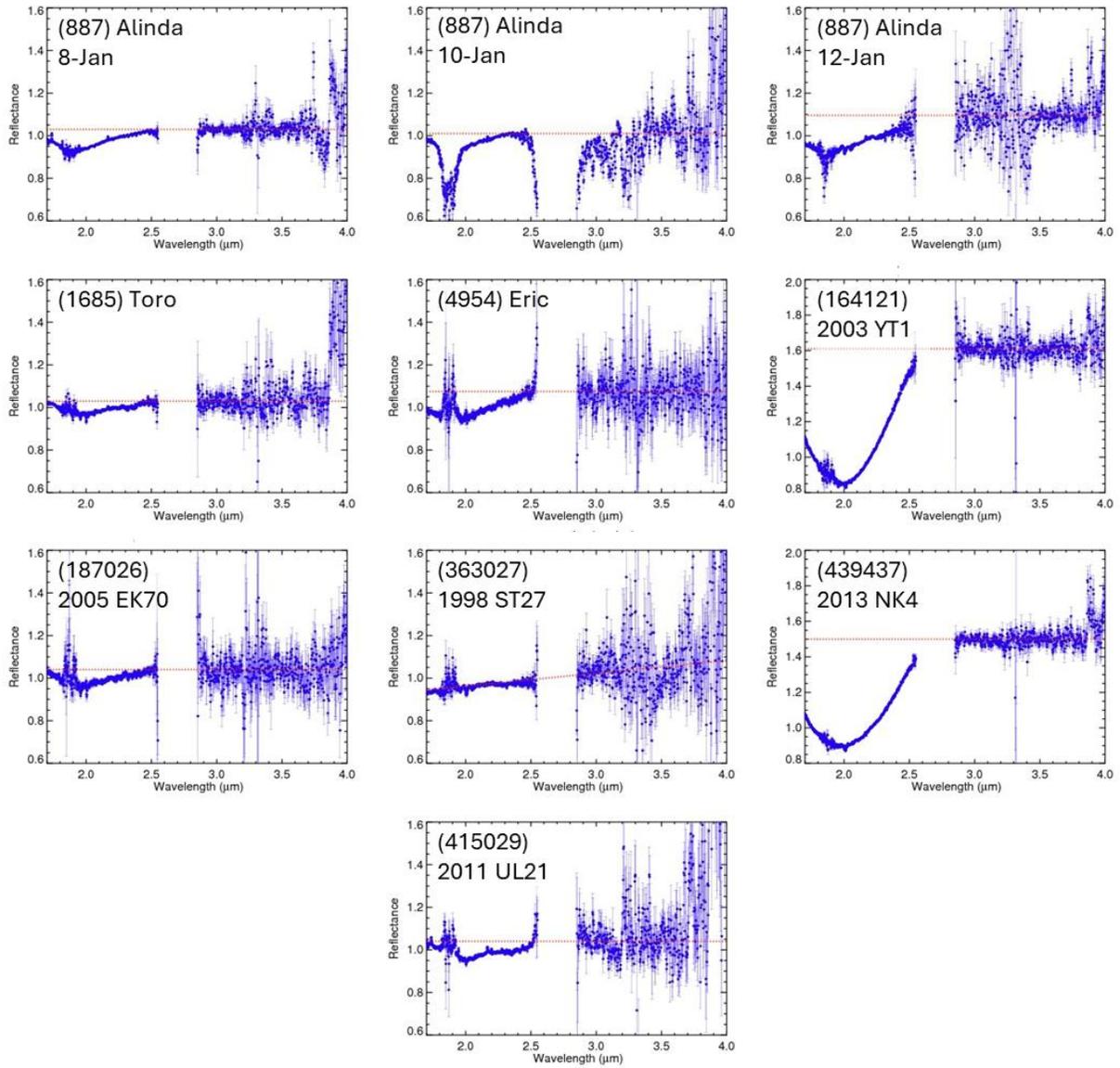

**Figure 2:** High S/N spectra of NEOs with no 3-μm absorption feature. The blue points represent the reflectance spectra (thermal component removed) and the red dashed lines are the reflected continua.

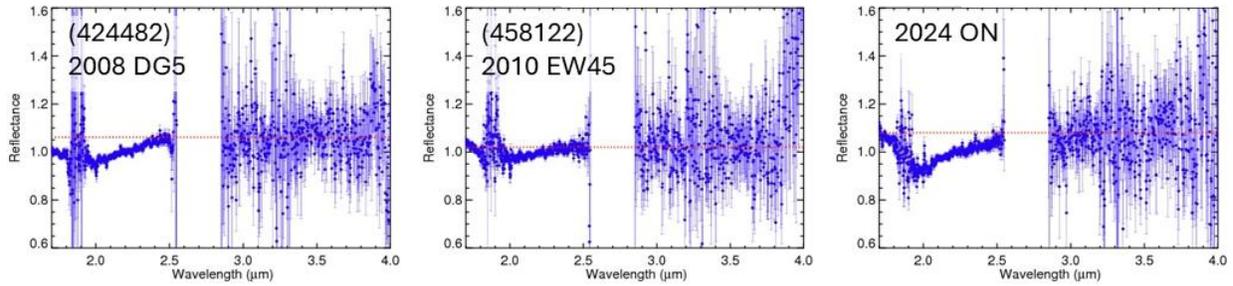

**Figure 3:** NEOs with low S/N spectra. The blue points represent the reflectance spectra (thermal component removed) and the red dashed lines are the reflected continua.

## 4 Analysis
### 4.1 Prism Band Parameter Analysis

Band parameters and calculated mineralogies are reported in Table 4 and BIC and BAR are plotted in Figure 4 overtop the Gaffey et al. (1993) S-complex asteroid sub-types for all S-complex and V-type NEOs observed in this study (Table 2). Figure 4 also includes the ordinary chondrite divisions defined by Dunn et al. (2010). All values reported and plotted have been corrected and adjusted for temperature, red- and blue-edges, and S/N (Burbine et al. 2007; Sanchez et al. 2020).

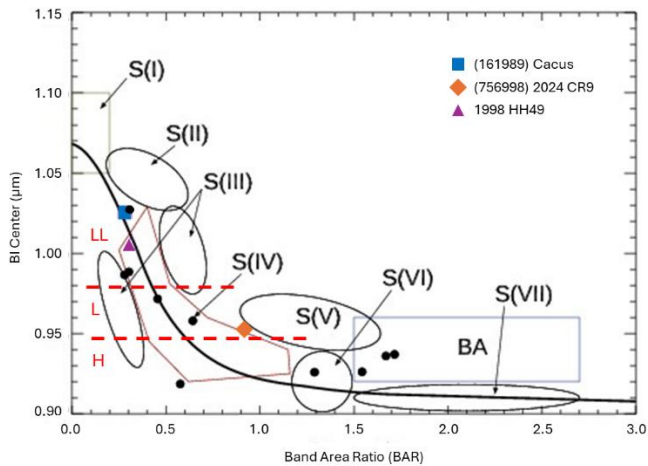

**Figure 4:** Gaffey et al. (1993) S-subtypes plot with ordinary chondrite lines from Dunn et al. (2010). The objects with (potential) 3-μm absorption bands are shown as colored symbols. The NEOs with no band are shown as black circles. Note that both detections and non-detections occur throughout the ordinary chondrite region (SIV), but the basaltic achondrite (BA) region only contains non-detections.

Twelve of the sixteen observed NEOs are spectrally in the S-complex and two are V-types. The V-types ((164121) 2003 YT1 & (439437) 2013 NK4) plot in the basaltic achondrite (BA) region, along with the S-type (887) Alinda's average. The values for each individual spectrum of Alinda have not been plotted. The prism spectrum collected for Alinda on 10 January plots in the H ordinary chondrite region of the S(IV) "boot;" the data collected on 8 and 12 January plot in the BA region. (415029) 2011 UL21 plots in the S(VI) region. The remaining ten NEOs plot in or near the S(IV) ordinary chondrite "boot". (4954) Eric plots just outside the bottom of the S(IV) H ordinary chondrite region. Three NEOs ((1685) Toro, (424482) 2008 DG5, & (756998) 2024 CR9) plot inside the S(IV) L ordinary chondrite region, with 2024 CR9 lying just outside the bounds of the "boot." Three NEOs ((187026) 2005 EK70, (458122) 2010 EW45, & 1998 HH49) plot in the S(IV) LL ordinary chondrite region, with the former two along the lower BAR S(IV)-S(III) boundary. Finally, two NEOs ((161989) Cacus & 2024 ON) plot along the olivine mixing line above the S(IV) region, with nearly identical values of BIC and BAR.

The Sanchez et al. (2020) equations used for the S-complex asteroids are only intended to be applied to asteroids that plot within the S(IV) "boot," and thus they may not be formally applicable to Alinda and 2011 UL21. However, utilizing more reliable indicators of mineralogy and ordinary chondrite groups such as ferrosilite mole percent (Fs) vs. fayalite mole percent (Fa; Dunn et al. 2010; Lucas et al. 2018) show that these two NEOs plot just outside the H ordinary chondrite region determined from meteorite analysis, but within that region when spectrally-derived errors are included (Figure 5). The Sanchez et al. (2020) equations are therefore still a reliable means to calculate the mineralogy for these two NEOs. 2024 CR9 and 1998 HH49 still plot in the L and LL ordinary chondrite regions, respectively. Cacus plots within the LL ordinary chondrite region. Toro is the only object studied with previously published band parameters (McGraw et al. 2022); our mineralogical values match within one sigma.

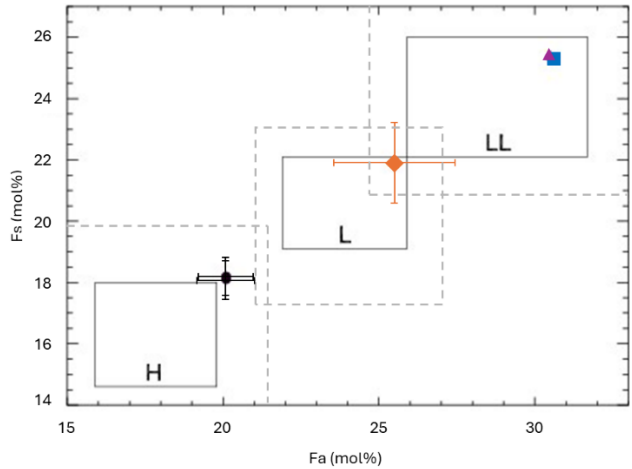

**Figure 5:** Modified ferrosilite vs. fayalite plot from Lucas et al. (2018) and Dunn et al. (2010) showing spectrally-derived ordinary chondrite sub-types. Solid boxes represent measured Fs and Fa ranges in ordinary chondrites from Brearley & Jones (1998). Dashed boxes represent least square root mean of the errors calculated by Dunn et al. (2010). (161989) Cacus is represented by a blue square, (756908) 2024 CR9 by an orange diamond, and 1998 HH49 by a purple triangle. The two overlapping black circles represent (415029) 2011 UL21 and the average value for (887) Alinda. Error bars for Cacus and 1998 HH49 are smaller than the plotted symbols.

**Table 4:** NEO band parameters

| S-Complex Near-Earth Objects | | | | | | |
|---|---|---|---|---|---|---|
| **Object** | **Fa** | **Fs** | **O/ol+pyx** | **BIC** | **BIIC** | **BAR** |
| 887 Alinda (8-Jan) | 20.2 ± 1.9 | 18.2 ± 1.3 | 0.359 ± 0.019 | 0.926 ± 0.008 | 2.005 ± 0.007 | 1.511 ± 0.080 |
| 887 Alinda (10-Jan) | 20.3 ± 1.8 | 18.2 ± 1.3 | 0.471 ± 0.016 | 0.927 ± 0.008 | 1.989 ± 0.007 | 1.039 ± 0.067 |
| 887 Alinda (12-Jan) | 19.9 ± 1.9 | 18.0 ± 1.3 | 0.215 ± 0.034 | 0.925 ± 0.008 | 2.015 ± 0.007 | 2.118 ± 0.142 |
| 1685 Toro | 28.1 ± 0.9 | 23.7 ± 0.6 | 0.610 ± 0.005 | 0.972 ± 0.008 | 2.016 ± 0.007 | 0.455 ± 0.019 |
| 4954 Eric* | 18.3 ± 2.0 | 16.9 ± 1.4 | 0.581 ± 0.028 | 0.919 ± 0.008 | 1.958 ± 0.022 | 0.576 ± 0.119 |
| 161989 Cacus | 30.6 ± 0.2 | 25.3 ± 0.2 | 0.651 ± 0.003 | 1.026 ± 0.008 | 2.011 ± 0.007 | 0.281 ± 0.011 |
| 187026 2005 EK70 | 29.6 ± 0.6 | 24.7 ± 0.4 | 0.646 ± 0.003 | 0.988 ± 0.008 | 2.015 ± 0.007 | 0.302 ± 0.014 |
| 415029 2011 UL21 | 20.1 ± 1.9 | 18.1 ± 1.3 | 0.411 ± 0.023 | 0.926 ± 0.008 | 1.987 ± 0.036 | 1.290 ± 0.096 |
| 424482 2008 DG5 | 26.2 ± 1.2 | 22.4 ± 0.8 | 0.565 ± 0.007 | 0.958 ± 0.008 | 1.995 ± 0.007 | 0.643 ± 0.029 |
| 458122 2010 EW45 | 29.5 ± 0.6 | 24.7 ± 0.4 | 0.652 ± 0.003 | 0.987 ± 0.008 | 2.013 ± 0.007 | 0.279 ± 0.011 |
| 756998 2024 CR9 | 25.5 ± 3.9 | 21.9 ± 2.7 | 0.500 ± 0.080 | 0.953 ± 0.023 | 1.892 ± 0.012 | 0.918 ± 0.335 |
| 1998 HH49 | 30.5 ± 0.2 | 25.4 ± 0.2 | 0.643 ± 0.003 | 1.007 ± 0.008 | 2.047 ± 0.007 | 0.315 ± 0.011 |
| 2024 ON | 30.5 ± 0.2 | 25.3 ± 0.2 | 0.646 ± 0.003 | 1.027 ± 0.008 | 2.002 ± 0.007 | 0.301 ± 0.012 |
| V-Type Near-Earth Objects | | | | | | |
| **Object** | **Fs** | **Wo** | **Mg #** | **BIC** | **BIIC** | **BAR** |
| 164121 2003 YT1 | 45.1 ± 8.2 | 10.6 ± 3.2 | 49.1 ± 10.3 | 0.937 ± 0.008 | 1.951 ± 0.007 | 1.716 ± 0.077 |
| 439437 2013 NK4 | 44.1 ± 8.2 | 10.2 ± 3.2 | 50.3 ± 10.3 | 0.936 ± 0.008 | 1.954 ± 0.007 | 1.670 ± 0.063 |

*Spectrum from online SMASS database

## 4.2 OH/$H_2O$ Band Analysis

Hydration features evinced by a 3-μm absorption band have been described by numerous studies on Main Belt Asteroids (MBAs), utilizing band shapes to determine exact volatile composition (e.g., water as opposed to hydroxide) and delivery mechanism (e.g., Rivkin 2010; Takir & Emery 2012; Rivkin et al. 2019; McAdam et al. 2024). Similar methods have been used to relate band shapes on NEOs to volatile composition and delivery mechanism (e.g., Rivkin et al. 2018; McGraw et al. 2022, 2024a, 2024b), though this population of asteroids has not been as well characterized. Linear or "sharp" bands have been linked to the presence of hydroxide (OH) in phyllosilicates or unbound molecules, which can be created via solar wind hydrogen implantation (e.g., Lebofsky 1980; Starukhina 2001, 2003, 2006; Farrell et al. 2015, 2017; Sunshine et al. 2009). Rounded or "bowl-like" bands have been linked to water ($H_2O$) or water ice, a mix of OH and $H_2O$, or even ammonia-bearing species (Takir and Emery 2012; De Sanctis et al. 2018; Poch et al. 2020). Of the eight NEOs McGraw et al. (2022) found to have 3-μm features, two possess linear bands and six possess rounded features (Table 5). That study broke those two groups into two sub-groups, narrow and wide, yielding four total band shape groups.

The four NEOs with (potential) bands described in this work all have band shapes that generally fit into the scheme described in McGraw et al. (2022; Figure 6). The NEOs identified in Section 3.1 exhibit each of the four band types summarized in Table 5. Cacus's feature most closely resembles that of a wide bowl (Figure 5A), similar to (1036) Ganymed and the Type 3 band. Its band is relatively flat from 2.9 μm to 3.2 μm, at which point it rapidly increases in reflectance to the reflected continuum. The feature exhibited by 2024 CR9's spectrum is a wide linear or Type 1 feature (Figure 5B), which is shared by (433) Eros. It linearly increases in reflectance from 2.9 μm to the reflected continuum at ~3.1 μm. 1998 HH49 possesses a narrow linear or Type 2 band (Figure 5C), similar to that of (214088) 2004 JN13; the reflectance increases linearly from 2.9 μm to ~3.05 μm. 2006 WB, though not an S-complex object like those described above or those with bands in McGraw et al. (2022), exhibits a band most similar to a narrow bowl (Type 4; Figure 5D), which was exhibited by five NEOs in McGraw et al. (2022). Type 4 bands have spectra that are generally flat from 2.9 μm to 2.95 μm, that then nonlinearly increase to the reflected continuum at ~3.0 μm; 2006 WB's band is shaped more like a "U," with the reflectance also meeting the continuum around 2.85 μm.

**Table 5:** NEO band shape descriptions

| Band Shape Type | Description | NEOs with Band Shape Type | |
|---|---|---|---|
| One | Wide linear: Reflectance linearly increases from ~2.9 μm to reflected continuum at ~3.1 μm | 756998 | 2024 CR9 |
| | | *433* | *Eros* |
| Two | Narrow linear: Reflectance linearly increases from ~2.9 μm to reflected continuum at ~3.05 μm | | 1998 HH49 |
| | | *214088* | *2004 JN13* |
| Three | Wide bowl: Feature shallow and flat from ~2.9 to ~3.1 μm, reflectance increases non-linearly to reflected continuum at ~3.2 μm | 161989 | Cacus |
| | | *1036* | *Ganymed* |
| Four | Narrow bowl: Feature shallow and flat from 2.8 to 3.0 μm, reflectance increases non-linearly to reflected continuum around 3.05 μm | | 2006 WB |
| | | *3122* | *Florence* |
| | | *96590* | *1998 XB* |
| | | *159402* | *1999 AP10* |
| | | *163373* | *2002 PZ39* |
| | | *671294* | *2014 JO25* |

*Note: NEAs in italics from McGraw et al. 2022. Table adapted from McGraw et al. 2022.*

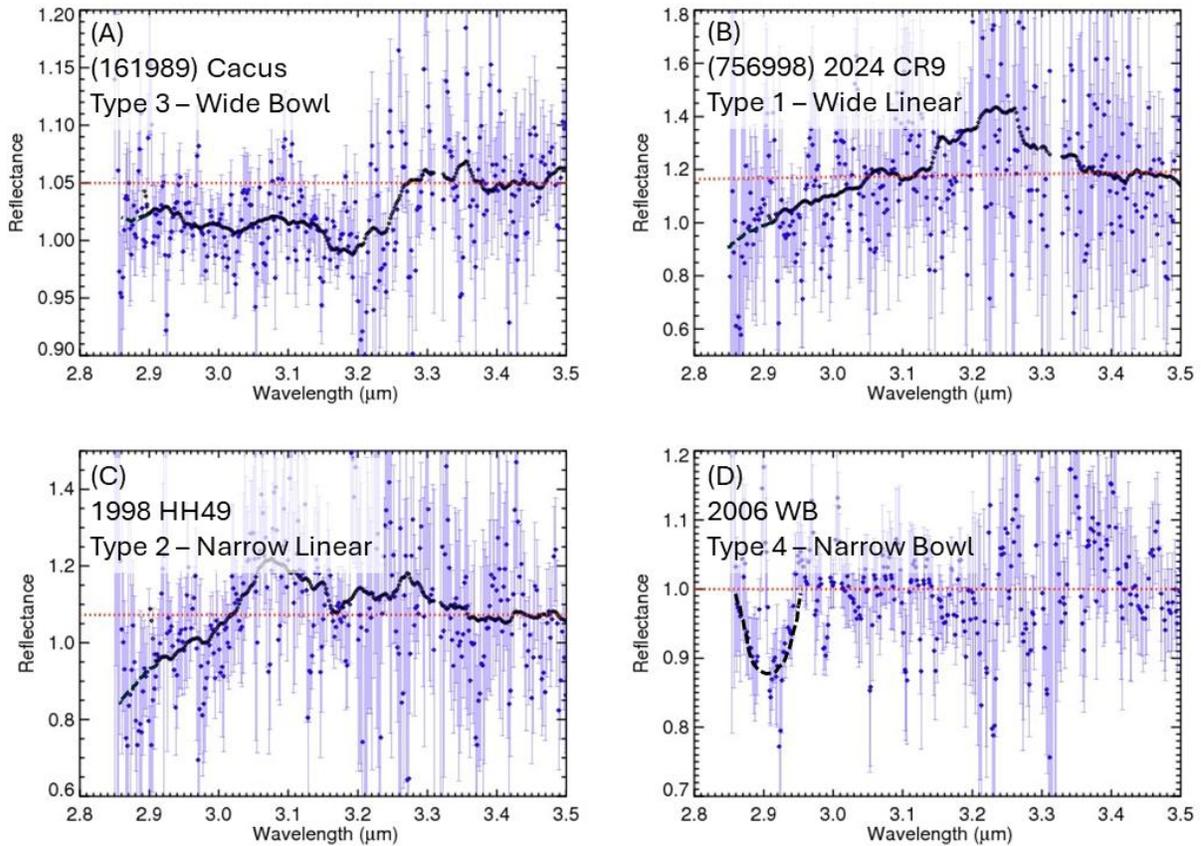

**Figure 6:** Spectra of NEOs with (potential) 3-μm absorption features, overlain by data filtered using a boxcar moving average with a width of 50. (A) Spectrum of (161989) Cacus as observed on 3&4 September 2022. (B) (756998) 2024 CR9 as observed on 9 June 2024. (C) 1998 HH49 as observed on 19 October 2023. (D) 2006 WB as observed on 27 November 2024. Its feature is too narrow to use the filter applied to the other spectra and so a manual estimation has been shown. The blue points represent the reflectance spectra (thermal component removed), the red dashed lines are the reflected continua, and the black solid lines represent the filtered data. Dashed black lines represent filter extrapolation.

## 4.3 Characteristics Comparisons

Numerous observational, physical, and orbital characteristics were compared against the 2.9-μm band depth to determine the factors controlling OH/$H_2O$ delivery and/or retention. Band depth was plotted against heliocentric and observer distances, perihelion, aphelion, orbital period, time since perihelion, Earth and Jupiter MOID, phase angle, inclination, eccentricity, rotation period, albedo, diameter, band I center (BIC), and band-area ratio (BAR). We also compared band depth against delta-V ($\Delta V$), which estimates the impulse per unit of spacecraft mass that is needed to rendezvous with the target, which in this case is each of our observed NEOs. NEOs with lower $\Delta V$'s are more easily accessible by spacecraft and are therefore more ideal as mission targets.

We added our dataset to that published by McGraw et al. (2022) in order to compare a larger population of targets. The most significant trend for NEOs discovered by McGraw et al. (2022) was that S-complex NEOs with aphelia in the Main Belt (Q > 2.06 AU) are more likely to possess a 3-μm feature. However, of the six high S/N S-complex NEOs with Q > 2.06 AU observed in this study (Table 2), only two have potential features (2024 CR9 and 1998 HH49). Further investigation yielded a slight trend in ΔV, the calculation of which depends on aphelion (Shoemaker & Helin 1978). As this is a derived quantity rather than one inherent to an NEO, we more closely analyzed the other variables used in its calculation and discovered a slight trend in orbital inclination as well (Figure 7). Plots of ΔV and orbital inclination show that band depth tends to increase with decreasing ΔV and inclination, and that most NEOs with (potential) bands in both datasets have ΔV < 10 km/s and i < 13.6°. One NEO has ΔV = 12.9 km/s ((671294) 2014 JO25), though i < 13.6°, and four have 13.6° < i < 26.7°.

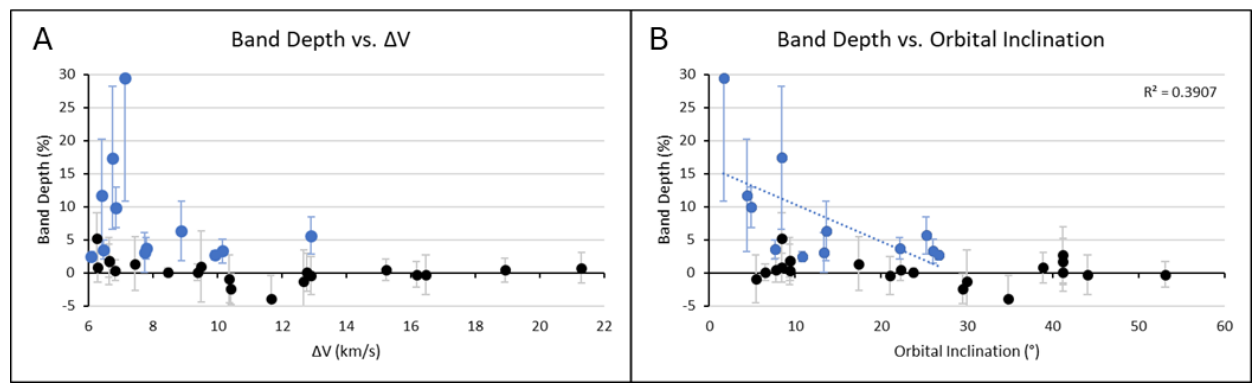

**Figure 7:** Plots showing band depth at 2.9 μm vs. ΔV (A) and orbital inclination (B) for all NEOs with high S/N spectra. Data from McGraw et al. 2022 and 2024b are included along with our newly collected data. Dotted lines represent the best-fit linear trend with $R^2$ value.

The only other potential trend of note is with BAR, as all three S-complex NEOs with bands we observed have BAR < 1. Six of the eight NEOs with bands from McGraw et al. (2022) also have BAR < 1, though the two remaining with BAR > 1 suggest BAR is not a strong driver of OH/$H_2O$ presence. Though no other trends in the sample as a whole were discovered, surface temperature and BIIC may play a role for some of our targets. Cacus exhibits the lowest surface temperature ($T_{sfc}$ = 234 K) of the NEOs we observed, though some NEOs in McGraw et al. (2022) possessed slightly lower temperatures at the time of observation. Conversely, 2024 CR9, 1998 HH49, and 2006 WB all have surface temperatures greater than 280 K. The only NEO observed by McGraw et al. (2022) with $T_{sfc}$ > 280 K is (163373) 2002 PZ39, which also has a potential 3-μm feature. For BIIC, 2024 CR9 has the lowest BIIC of all NEOs compared. 1998 HH49 has the highest BIIC of the NEOs observed in this study; three NEOs observed by McGraw et al. (2022) exhibited higher BIIC though only one of those also exhibited a potential 3-μm band ((96590) 1998 XB).

## 5 Discussion

The fifteen near-Earth objects (NEOs) studied in this survey represent a marked increase in the overall population of NEOs with spectral data in the 2-4-μm region. Only one has been previously studied with the intention of characterizing OH/$H_2O$ on its surface ((1685) Toro;

McGraw et al. 2022). Combining our data with previously published results has provided new understanding on how OH/$H_2O$ is delivered and/or retained.

## 5.1 Band Shape

As previously mentioned, McGraw et al.'s (2022) 3-µm NEO survey defined four band shape types (Table 5). Each of the four NEOs from this work with (potential) hydration features exhibits one of those same bands. However, the band shapes evident with our limited spectra after removing the wavelengths lost due to telluric water (2.45-2.85 µm) may not be the true shape of the band. Ground-based spectra of Psyche suggested a linear feature indicative of OH, but spectra from JWST showed Psyche possesses a rounder hydration feature, indicative of $H_2O$ (Jarmak et al. 2024). The following discussion is based solely on the part of the band visible from ground-based observations.

### 5.1.1 (756998) 2024 CR9 – Wide Linear

NEO (756998) 2024 CR9 was observed on 9 June and 8 August 2024 in LXD_short and prism modes, respectively (Figures 1B & 5B). Analysis of the prism spectrum shows that this NEO is an S-type asteroid that most closely matches L-type ordinary chondrite meteorites (Table 2; Figures 4 & 5). It has an aphelion of 3.65 AU and was likely sourced from the $\nu_6$ secular resonance region (74% probability) in the Main Belt (Granvik et al. 2018; Moskovitz et al. 2022).

Its band depth of 11.7 ± 8.6% marks it as a "potential" band, as the band depth is deeper than the 1σ error but not the 2σ. 2024 CR9's band shape is best described as wide and linear, as seen on (433) Eros by McGraw et al. (2022). Eros' band shape was defined as "Band Type 1" in that study. Other objects with similarly shaped 3-µm bands have been described on Main Belt Asteroids (MBAs) in Takir & Emery (Pallas-like; 2012), Rivkin et al. ("sharp"; 2019), and McAdam et al. (2024). This band shape is also reminiscent in shape to that of the global lunar hydration feature (e.g., Sunshine et al. 2009; Pieters et al. 2009; Clark 2009; fig. 12a from McGraw et al. 2024a). 2024 CR9's surficial OH/$H_2O$ is therefore most likely dominated by OH, rather than $H_2O$, and may have been delivered via solar wind hydrogen implantation. This volatile delivery mechanism is theorized to have delivered OH to the Moon (e.g., Sunshine et al. 2009; Wöhler et al. 2017; Laferriere et al. 2022) and Eros (McGraw et al. 2022; McGraw et al. 2024a). Though this band is potentially deep enough to suggest the OH is bound within a mineral structure, native phyllosilicates are an unlikely hydration source on an S-type asteroid.

### 5.1.2 1998 HH49 – Narrow Linear

1998 HH49 was observed on 19 October 2023 (Figure 1C). It has an aphelion of 2.33 AU (Table 2) and was also likely sourced from the $\nu_6$ secular resonance region (86% probability) in the Main Belt (Granvik et al. 2018; Moskovitz et al. 2022). Analysis of our prism spectrum places it as a Q-type asteroid that would meteoritically classify as an LL-ordinary chondrite (Figures 4 & 5).

1998 HH49 has the deepest band at 17.4 ± 10.9% and is considered a "potential" band (Figures 1C & 5C). 1998 HH49's 3-µm band shape is similar to that of 2024 CR9, but narrower, increasing in reflectance to the reflected continuum around 3.05 µm, ~0.15 µm shorter than for 2024 CR9. 1998 HH49's band is therefore classified as a narrow linear band, which was defined as a Type 2 band in McGraw et al. (2022). Only one NEO in that study, (214088) 2004 JN13, exhibited this band shape, which was attributed to OH rather than $H_2O$. We similarly conclude that 1998 HH49's 3-μm band shape is predominantly caused by OH, likely delivered via solar wind hydrogen implantation.

As with 2024 CR9, 1998 HH49's band depth could suggest a native mineralogical hydration source, but this is unlikely on a Q-type asteroid, which is considered to be a "fresher" or less space-weathered version of an S-type asteroid (McFadden et al. 1985; Binzel et al. 2010). Without the complete hydration band, it is unclear what might cause the different band widths exhibited by 2024 CR9 and 1998 HH49.

### 5.1.3 (161989) Cacus – Wide Bowl

(161989) Cacus was observed on 3 & 4 September 2022 and is the largest of the four NEOs with (potential) features (D = 1.9 km; Table 2; Figures 1A & 5A). Its aphelion means Cacus no longer interacts with the Main Belt (Q = 1.36 AU) and was also likely sourced from the $\nu_6$ secular resonance (62% probability) or possibly the Hungaria (28% probability) region in the Main Belt (Granvik et al. 2018; Moskovitz et al. 2022). Cacus is the only NEO of the four with (potential) features that had been observed in prism prior to this study, and our spectral classification of this object as a Q-type is consistent. It plots as an LL-ordinary chondrite meteorite (Figures 4 & 5).

Cacus has the shallowest band depth of the four NEOs of 3.3 ± 2.4%. Its band shape is best described as a wide bowl, very similar to McGraw et al.'s (2022) Type 3 band exhibited by (1036) Ganymed. McGraw et al. (2022; 2024a) showed that Ganymed's band shape is very similar to that of (4) Vesta, the largest V-type MBA, and (349) Dembowska, a large R-type MBA. Vesta was thoroughly studied by NASA's Dawn mission (Russell et al. 2012), during which low-albedo regions caused by exogenous carbonaceous material impacts were linked to regions with hydration absorption features centered at ~2.7 μm (e.g., De Sanctis et al. 2012; Reddy et al. 2012a; McCord et al. 2012). Cacus, like Ganymed, may therefore possess exogenous carbonaceous material.

Additionally, rounded features have been linked to the presence of $H_2O$ (Takir & Emery 2012), as well as ammonia-bearing clays and salts (e.g., De Sanctis et al. 2018; Poch et al. 2020). Ganymed's aphelion of 4.09 AU means it regularly interacts with the outer Main Belt, a region with significantly more carbonaceous material than the region of space occupied by Cacus (DeMeo et al. 2015). It is unlikely that Cacus interacted with carbonaceous material, though not impossible, given its short aphelion and likely inner Main Belt source region (the $\nu_6$ secular resonance region is at ~2.06 AU).

### 5.1.4 2006 WB – Narrow Bowl

2006 WB was observed on 27 November 2024 and, unlike the NEOs discussed in Sections 5.1.1-5.1.3, is not in the S-complex (Table 2; Figures 1D & 5D). No other ground-based NEO 3-μm survey has confirmed hydration absorption features in the 2-4-μm region for non-S-complex objects (e.g., McGraw et al. 2022), though ground- and space-based spectra of hydrated non-S-complex NEOs do exist (e.g., Volquardsen et al. 2007; Rivkin et al. 2013; Kitazato et al. 2019; Hamilton et al. 2019). The online Bus-DeMeo classification tool (DeMeo et al. 2009) suggests that 2006 WB is either an X- or C-complex asteroid; the tool was unable to differentiate between these two complexes without shorter wavelength data. 2006 WB is also the smallest of the four NEOs with (potential) 3-μm features (D = 0.1 km) and has the shortest aphelion (Q = 1.00 AU). Like the other three NEOs with (potential) hydration features, it was likely sourced from the $\nu_6$ secular resonance region (82% probability) in the Main Belt (Granvik et al. 2018; Moskovitz et al. 2022).

It is the only NEO with a definitive band, as its band depth of $9.9 \pm 3.4\%$ is deeper than the $2\sigma$ error. Its band shape most closely resembles McGraw et al.'s (2022) Band Shape Type 4, a narrow bowl, though it is not an exact match. This feature is similar to that of Cacus', but narrower. This band shape type was seen by five of the eight NEOs in McGraw et al. (2022) with (potential) 3-$\mu$m absorption features.

2006 WB is significantly different from the other NEOs with Band Shape Type 4 observed by McGraw et al. (2022), as it is not in the S-complex and does not interact with the Main Belt. The similar band shapes still suggest a similar volatile composition, such as $H_2O$ ice in addition to OH (Takir & Emery 2012; McGraw et al. 2022, 2024a), but the delivery mechanism is less easily constrained. The bowl-like feature generally removes solar wind hydrogen implantation as a source, as this mechanism should more easily create OH rather than $H_2O$ (e.g., Starukhina 2001, 2003, 2006; Farrell et al. 2015, 2017) unless some other mechanism such as recombinative desorption causes the resultant OH molecules to bind together into $H_2O$ (Orlando et al. 2018). As for exogeneous carbonaceous material, 2006 WB's orbital parameters and probable source region indicate a low probability of carbonaceous impactors. Furthermore, 2006 WB's inherent mineralogy means that native phyllosilicates could be considered as a hydration source, except bands indicative of native phyllosilicates or those created via aqueous alteration are typically linear in shape (e.g., Rivkin et al. 1995, 2000; Landsman et al. 2015; Takir et al. 2017; Hamilton et al. 2019; Nakamura et al. 2023; Zega et al. 2025; Pilorget et al. 2022, 2025). It is possible that 2006 WB's inherent surface mineralogy interacts with the solar wind in a different manner than nominally anhydrous asteroids, or that relatively high surface temperatures due to its low perihelion (q = 0.77 AU) and aphelion have altered any nominally hydrous materials present at formation. Analysis of other non-S-complex NEOs with 3-$\mu$m features is needed to further determine the OH/$H_2O$ delivery mechanism(s) active on 2006 WB.

## 5.2 Potential Driving Factors

### 5.2.1 Inclination

Previous work showed S-complex NEOs with aphelia in the Main Belt are more likely to exhibit a 3-$\mu$m feature (McGraw et al. 2022), but our data suggests this trend is not as strong as anticipated (Figure 8). However, we do see some relation between band presence and depth with $\Delta V$ (Figure 7), the calculation of which depends on many factors including aphelion (Shoemaker & Helin 1978). Another variable on which these $\Delta V$ equations depend is orbital inclination. We investigated all quantities in the $\Delta V$ equations for trends but found that inclination was the only one with a noticeable relationship to band presence and depth. Inclination was tested for trends in previous surveys (e.g., McGraw et al. 2022) but was not found to be of any importance. Combining our new dataset with that of McGraw et al. (2022) revealed that all NEOs with a band have i < 27° and that two-thirds have i < 14° (Figure 7). Additionally, band depth increases as inclination decreases. These promising new relationships show the importance of continuing to add new NEO 3-$\mu$m spectra to this database, as the trends in inclination were only evident with MITHNEOS' additional spectra. The trend seen by McGraw et al. (2022) between band presence and aphelion was likely hinting at a more complicated relationship between NEO orbit and band presence, which may help explain why some low-aphelion NEOs exhibit a hydration feature.

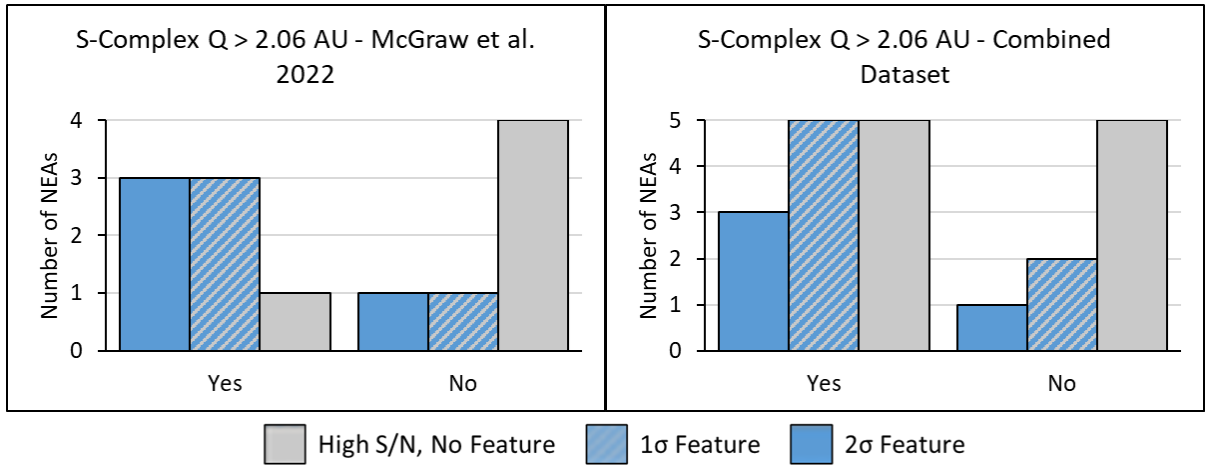

**Figure 8:** (Left) Histogram of higher S/N S-complex NEOs observed by McGraw et al. 2022 (see fig. 12). (Right) Histogram of higher S/N S-complex NEOs observed by McGraw et al. 2022, McGraw et al. 2024b, and this work. Solid blue represents objects with 2.9-μm bands detected to $\geq 2\sigma$ and hashed blue are the objects with 1-2$\sigma$ detections. Solid gray represent the NEOs with no detections and upper limits $\leq 5\%$.

Though intriguing, the trend in inclination is non-exclusive. Several NEOs that also have low inclinations do not have 3-μm features. Analysis of variance (ANOVA) statistical tests of the combined MITHNEOS-McGraw et al. (2022) dataset show that the difference in the mean inclination of the NEOs with (potential) features ($\bar{i} = 13.7°$) and the mean inclination of all high S/N NEOs without a band ($\bar{i} = 22.9°$) has a 6% probability of being statistically the same, or in other words the means of the two populations have a high probability of being statistically different. However, when comparing S-complex NEOs with (potential) features ($\bar{i} = 14.5°$) and high S/N S-complex NEOs with no features ($\bar{i} = 18.0°$), the probability of the difference being statistically significant is nearly the same as the probability that the different mean inclinations are statistically identical. Removing V-type and X/C-complex NEOs drastically reduces the likelihood that the inclination trend in band presence is statistically significant and suggests band presence may be more dependent on spectral type than inclination.

To date, the only V-type asteroid with a confirmed hydration feature is the Main Belt asteroid (4) Vesta (e.g., De Sanctis et al. 2012), the archetypal V-type asteroid. McGraw et al. (2022) hypothesized that V-types, composed primarily of pyroxene rather than a mixture of pyroxene and olivine like the S-complex, are less susceptible to solar wind hydrogen implantation due to the stronger molecular bonds in pyroxene compared to olivine. However, six of the seven V-type NEOs observed by this study and McGraw et al. (2022) have i > 14° and five have i > 27°. V-type NEO inclinations are widely variable, though the larger V-types have higher inclinations (Galiazzo et al. 2016) and these larger V-types have been preferentially observed due to more ideal visible magnitudes. It is possible that the lack of 3-μm features on observed V-types is due to their high inclination rather than their inherent mineralogy, but observing smaller V-type NEOs with low inclinations is necessary to determine the role of composition versus inclination for these bodies. However, as (439437) 2013 NK4 is a low-inclination V-type and does not have a 3-μm feature (Tables 2 & 3), we predict composition to be more important than inclination for the lack of surficial OH/$H_2O$ on V-type NEOs.

Orbital inclination may be critical to OH/H$_2$O presence on NEOs for two reasons. One, objects with lower inclinations are by definition closer to the ecliptic, as are other planetary bodies such as the major planets, MBAs, and other NEOs. The Solar System in general possesses more material closer to the ecliptic, so NEOs with lower inclinations are more likely to be impacted by exogeneous material, carbonaceous or otherwise, than NEOs with higher inclinations. Two, during quiet solar periods, solar wind density is significantly higher in a 20-30° wedge along the plane of the Sun's equator, which is inclined to the ecliptic by roughly 7° (e.g., Richardson 2001). As solar activity increases, the solar wind density profile with latitude becomes more complex[2]. During the ~11-year solar cycle, NEOs closer to the plane of the solar equator, i.e., with inclinations closer to 7°, experience a greater and steadier flux of protons from the solar wind than NEOs with higher inclination, increasing the possibility of OH/H$_2$O formed by solar wind hydrogen implantation (e.g., Lebofsky 1980; Starukhina 2001, 2003, 2006; Farrell et al. 2015, 2017). The four NEOs with linear features likely indicative of OH (Eros, 2024 CR9, 2004 JN13, & 1998 HH49) have inclinations ranging between 4.4° and 13.3°, so have orbits generally aligned with the solar equatorial plane (Table 2). In either case, lower orbital inclination should increase the likelihood of surficial OH/H$_2$O regardless of delivery mechanism. Though the average inclination of NEOs is roughly similar to the highest inclinations seen by NEOs with hydration features ($\bar{i} \approx 20°$; Granvik et al. 2018), using this inclination cutoff will significantly narrow the search for hydration on NEOs.

### 5.2.2 Mineralogy

Nine of the eleven S-complex NEOs with (potential) 3-$\mu$m features have BAR values less than 1 (Table 4; table 4 in McGraw et al. 2022). As BAR is used to calculate the olivine ratio (i.e., ol/[ol+pyx]), this means the majority of S-complex NEOs with surficial OH/H$_2$O are composed of more than 48% olivine relative to the combined quantity of olivine and pyroxene. McGraw et al. (2022) posited that because the degree of silica saturation affects bond strength, asteroids that are composed of minerals with higher degrees of silica saturation such as pyroxene are resistant to solar wind hydrogen implantation because the stronger molecular bonds are less likely to be mechanically broken by the solar wind, which is the first step in creating OH/H$_2$O with this delivery mechanism (e.g., Farrell et al. 2015, 2017). Similarly, S-complex NEOs with lower BAR values are richer in less silica-saturated olivine and therefore may be more susceptible to OH/H$_2$O creation via solar wind hydrogen implantation due to their prevalence of weaker molecular bonds. Laboratory studies regarding solar wind hydration implantation have primarily focused on this mechanism's ability to form OH/H$_2$O, so have not detected any mineralogical trends in hydration band presence relative to olivine ratio (e.g., Burke et al. 2011; Bradley et al. 2014; Schaible & Baragiola 2014; Zhu et al. 2019; McClain et al. 2021). Band II center is not used in mineralogical calculations, so it is unclear how the low and high values of BIIC for 2024 CR9 and 1998 HH49, respectively, may affect OH/H$_2$O delivery and/or retention.

### 5.2.3 Temperature

The importance of surface temperature on the presence of a 3-$\mu$m absorption feature has been investigated in previous NEO hydration characterization studies (e.g., Rivkin et al. 2018; McGraw et al. 2022). Surface temperature has also been linked to the depth of this spectral feature on the Moon, with higher temperatures associated with shallower bands (McCord et al. 2011), though continued analysis of lunar data and this correlation is ongoing (e.g., Ruiz et al. 2020; Honniball et al. 2021; Chauhan et al. 2021). Arguments can be made for both low and high surface temperatures effects on OH/$H_2O$ presence. Lower surface temperatures, such as seen on Cacus ($T_{sfc} \approx 230$ K), create a more stable environment for OH/$H_2O$ retention. Conversely, higher temperatures, as seen on 2024 CR9, 1998 HH49, and 2006 WB, as well as (163373) 2002 PZ39 (McGraw et al. 2022), can aid processes such as recombinative desorption (Orlando et al. 2018). As discussed in Section 5.1.4, recombinative desorption can convert OH into $H_2O$ and was hypothesized by McGraw et al. (2022) to be involved in creating $H_2O$ on the NEOs with narrow bowl (Type 4) shaped bands. Rivkin et al. (2018) saw a slight trend in heliocentric distance, and therefore surface temperature at the time of observation, with band depth on (433) Eros and (1036) Ganymed, though subsequent work on those same targets could not confirm this trend (McGraw et al. 2024a). Though temperature seems a likely factor to influence the delivery and/or retention of volatiles such as OH and $H_2O$, a definitive connection has yet to be found.

---

[2] https://www.swpc.noaa.gov/phenomena/solar-wind#:~:text=Different%20regions%20on%20the%20Sun,densities%20and%20strong%20magnetic%20fields

## 5.3 Non-Detections

Understanding why some of our targets do not exhibit a detectable 3-μm feature can be just as useful in determining what controls OH/$H_2O$ delivery and/or retention as studying the NEOs with a 3-μm absorption feature. As discussed in Sections 5.2.1 and 5.2.2, we do not see a 3-μm feature on the two V-type NEOs we observed, bringing the total count of V-type NEOs with 2-4-μm spectra to seven (McGraw et al. 2022). Whether this is due to their pyroxene-rich composition or their high inclinations is as yet unclear, but it seems reasonable to predict that future observations of V-type NEOs will also result in spectra with no measurable 3-μm feature.

According to McGraw et al. (2022), S-complex NEOs that enter the Main Belt (Q > 2.06 AU), are more likely to possess surficial OH/$H_2O$ yet we observed three S-complex NEOs with sufficiently high S/N and Q > 2.06 AU that do not possess a 3-μm feature. (415029) 2011 UL21 and (887) Alinda, which will be discussed further in Section 5.4, have BAR > 1, so have lower olivine ratios and therefore relatively more pyroxene than the NEOs with 3-μm bands. If V-types do not possess surficial OH/$H_2O$ because they are primarily composed of pyroxene, perhaps Alinda and 2011 UL21 are similarly too pyroxene-rich to allow OH/$H_2O$ delivery/retention. However, (1036) Ganymed also has BAR > 1 and is well-documented to exhibit a 3-μm band (e.g., Rivkin et al. 2018; McGraw et al. 2022, 2024a). Its band shape suggests exogeneous carbonaceous material is the source of its hydration, so perhaps Alinda and 2011 UL21 simply have not yet been impacted by such material. Additionally, 2011 UL21 has a fairly high inclination (i = 34.9°), which may also make OH/$H_2O$ delivery/retention more difficult (see Section 5.2.1).

The final NEO in this study with a large aphelion but no hydration band is (4954) Eric. Its spectrum has S/N on the border between "high" S/N and "low" S/N, so it is possible it has a band with depth less than 4%. Alternatively, its size and orbit are similar to that of (1627) Ivar, another S-complex NEO predicted to possess surficial OH/$H_2O$ that did not (McGraw et al. 2022). Another S-complex NEO with large aphelion (i.e., Q > 2.06 AU) and low inclination (i.e., i < 27°) that does not exhibit a 3-μm feature is 2024 MK (McGraw et al. 2024b). Perhaps some as yet undetermined characteristic might explain the lack of 3-μm features on these three objects.

## 5.4 (887) Alinda Rotational Study

In January 2025, (887) Alinda reached a visible magnitude of 9.4 and was brighter than tenth magnitude for two weeks. The spin pole and orbital inclination of Alinda enabled us to schedule four observations centered roughly 90° apart to enable a rotationally resolved study. Unfortunately, the observation centered at 90° longitude relative to the first observation was cancelled due to fog and the observation centered at 270° longitude occurred while the skies were overcast for much of the observation window. The spectrum from 10 January 2025 (long. = 270°) is of lower quality and S/N than those taken on 8 & 12 January 2025 (long. = 0° & 180°, respectively; Figure 2).

The spectra collected at 0° and 180° relative longitude have band depths of 0.4 ± 1.9% and 0.9 ± 5.9%, respectively, indicating that the majority of Alinda's surface does not possess $OH/H_2O$. Each spectrum is disc-resolved and roughly covers half of Alinda's surface, so the spectra collected on 8 & 12 January 2025 sample roughly the entire surface. The band depth taken on 10 January 2025 at 270° relative longitude is 12.7 ± 3.8%. Considering that neither the 8 January nor 12 January spectra show any indication of $OH/H_2O$ presence, even though these three spectra cover overlapping portions of the surface, the deep band in the 10 January spectrum is likely due to poor telluric correction.

We also collected prism data at each longitude, though the frames from 10 January had widely varying slopes so we will again focus the discussion on 8 & 12 January. The two hemispheres are more elementally similar than mineralogically, as their band I centers (BIC) agree to within 1σ but their band-area ratios (BAR) are significantly different (Table 4). BIC is used to calculate the percentage of iron-rich olivine and pyroxene endmembers versus their magnesium-rich counterparts. The comparable BIC values across Alinda's surface indicate a similar iron-magnesium ratio. However, as BAR is used to calculate the percentage of olivine compared to the total amount of olivine and pyroxene, the different BAR values suggest that the hemisphere first observed (long. = 0°; BAR = 1.51 ± 0.08) has ~65% more olivine relative to the second hemisphere (long. = 270°; BAR = 2.12 ± 0.14). The BAR calculated for Alinda's 12 January spectrum is the largest by far of this survey even when V-types are included; this spectrum is noisier than the 8 January spectrum due to less exposure time which may explain the abnormally large BAR. Regardless, the results of our rotational study show Alinda lacks a hydration feature and is elementally, if not mineralogically, homogeneous.

## 6 Summary and Conclusions

We observed 15 near-Earth objects using near-infrared spectroscopy to continue characterizing $OH/H_2O$ in near-Earth space. By using NASA's IRTF Spex instrument to collect prism (0.7-2.5 μm) and LXD (1.67-4.2 μm) data, we tested published hypotheses regarding $OH/H_2O$ delivery and/or retention and investigated new trends. Four NEOs exhibit a 3-μm absorption feature indicative of surficial $OH/H_2O$ to 1- or 2σ: (161989) Cacus, (756998) 2024 CR9, 1998 HH49, and 2006 WB. Eight of the remaining eleven NEOs possess spectra of high enough S/N to determine the lack of a hydration feature to within a few percent band depth.

The four NEOs with 3-μm features possess band shapes similar to those previously published for nominally anhydrous NEOs, even though one (2006 WB) is not silicaceous. 2024 CR9 and 1998 HH49 exhibit wide and narrow linear features, respectively, suggesting hydroxide as the dominant hydrated material. Though these bands are deep enough to be caused by native phyllosilicates, such a source would be unexpected on S-complex asteroids. The predominant hypothesis explaining the presence of OH on nominally anhydrous asteroids is delivery via solar wind hydrogen implantation, which is also hypothesized for the global lunar hydration feature. Cacus and 2006 WB exhibit wide and narrow bowl-like features, respectively, which likely indicate the presence of water, possibly in addition to hydroxide. Other inner Solar System bodies with wide bowl-like features likely possess exogeneous carbonaceous material, such as on (4) Vesta. However, as Cacus' orbit does not carry it through regions of higher carbonaceous material concentration ($Q = 1.36$ AU), it is unlikely to have such material on its surface. 2006 WB possesses the most common 3-$\mu$m band shape seen on NEOs but is the only one not in the S-complex to exhibit it. This shape has been linked to multiple sources such as exogeneous carbonaceous material and/or solar wind hydrogen implantation combined with recombinative desorption, but native mineralogy may also play a role for this particular object. Observing these objects with a space telescope such as the James Webb Space Telescope or the in-development Twinkle mission (Edwards et al. 2019) would enable full-band studies that may further clarify their source(s) of surficial OH/$H_2O$.

Combining our data with previously published datasets in this spectral region, we note that band depth tends to increase with decreasing orbital inclination. Additionally, all NEOs with (potential) features have $i < 27°$. Lower inclination may aid in OH/$H_2O$ delivery because the solar wind density is often increased closer to the ecliptic and impacts with hydrated material are more likely. We also investigated the prediction that S-complex NEOs that enter the Main Belt are more likely to exhibit a 3-$\mu$m feature. We found that while this prediction still holds true, the probability that any S-complex NEO with $Q > 2.06$ AU will exhibit a 3-$\mu$m feature is significantly less than previously suggested. Additionally, we determined that NEOs with more olivine relative to pyroxene (i.e., BAR < 1) are more likely to exhibit a 3-$\mu$m feature. Increasing the total number of NEOs with 2-4-$\mu$m spectroscopy has enabled old hypotheses to be tested and new ones to emerge. This updated dataset allows us to predict that S-complex NEOs with low inclinations and large aphelia are the most likely to contain surficial OH/$H_2O$.

## Acknowledgements

Visiting Astronomer at the Infrared Telescope Facility, which is operated by the University of Hawaii under contract 80HQTR19D0030 with the National Aeronautics and Space Administration. The prism spectrum of (4954) Eric utilized in this publication was obtained and made available by the MITHNEOS MIT-Hawaii Near-Earth Object Spectroscopic Survey, the MIT component of which is supported by NASA grant 80NSSC18K0849. This work was supported by NASA's Yearly Opportunities for Planetary Defense (YORPD) grant 80NSSC22K0773.

## Appendix A

## S-Complex

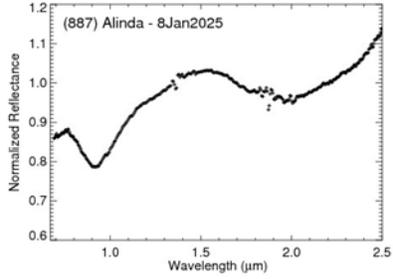 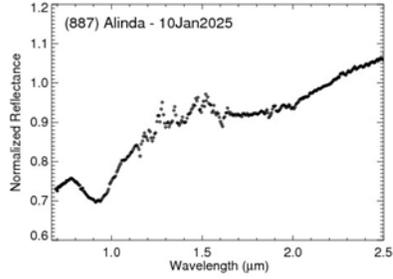 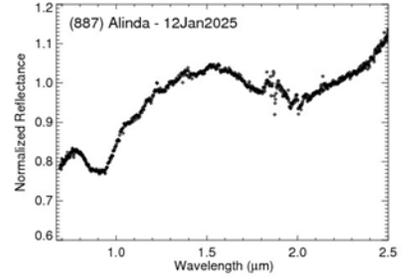

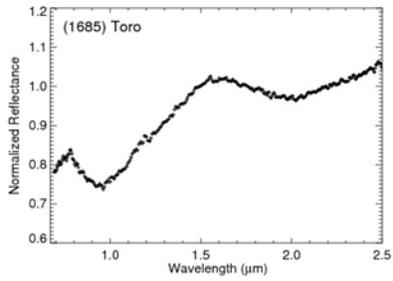 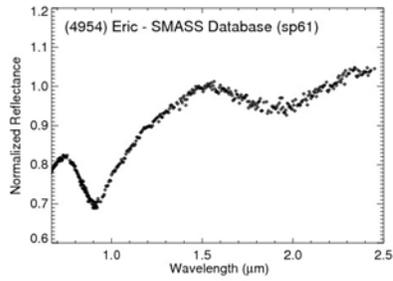 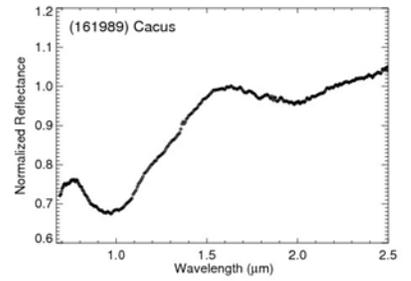

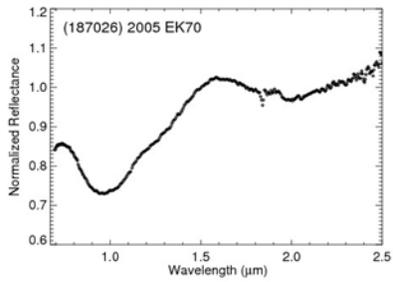 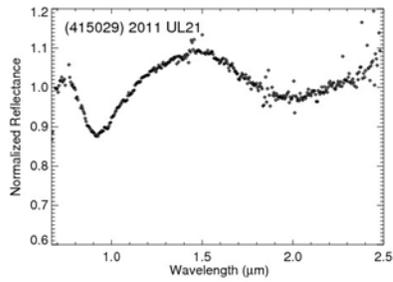 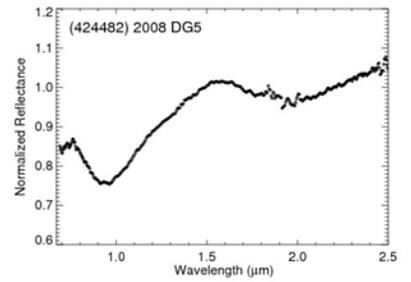

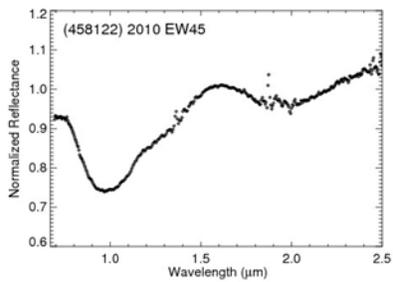 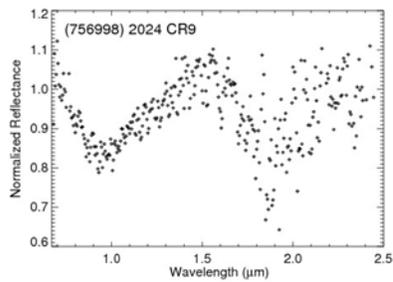 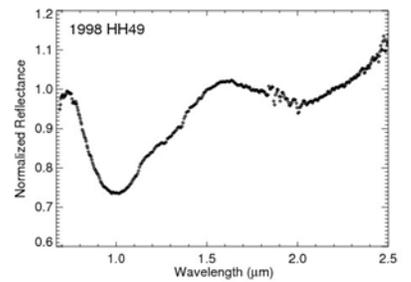

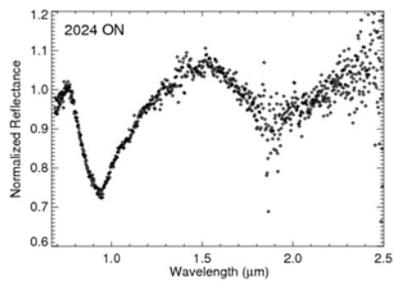

## X/C-Complex

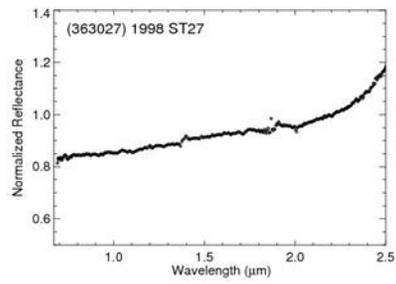 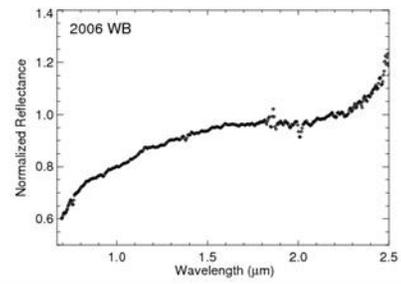

## V-Types

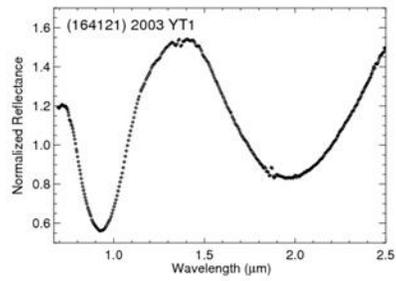 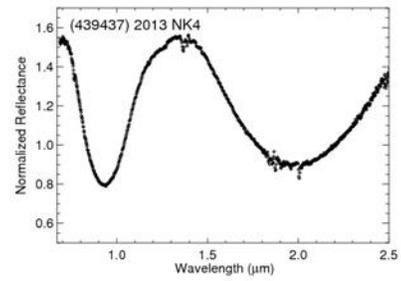